\documentclass[11pt]{article}
\usepackage{moriond,epsfig}
\pdfoutput=1 

\def\be{\begin{equation}}
\def\ee{\end{equation}}
\def\bea{\begin{eqnarray}}
\def\eea{\end{eqnarray}}

\begin{document}
\vspace*{4cm}
\title{Thermal Dimuon Emission in In-In at the CERN SPS}
\author{
  R.~Arnaldi$^{10}$,
  K.~Banicz$^{3,5}$,
  J.~Castor$^{4}$,
  B.~Chaurand$^{7}$,
  C.~Cical\`o$^{2}$,
  A.~Colla$^{9,10}$,
  P.~Cortese$^{9,10}$,
  S.~Damjanovic$^{3,5}$,
  A.~David$^{6}$,
  A.~de~Falco$^{1,2}$,
  A.~Devaux$^{4}$,
  L.~Ducroux$^{11}$,
  H.~En'yo$^{8}$,
  J.~Farjeix$^{4}$
  A.~Ferretti$^{9,10}$,
  M.~Floris$^{1,2}$,
  A.~F\"orster$^{3}$,
  P.~Force$^{4}$,
  N.~Guettet$^{3,4}$,
  A.~Guichard$^{11}$,
  H.~Gulkanyan$^{12}$,
  J.~Heuser$^{8}$,
  M.~Keil$^{3,6}$,
  L.~Kluberg$^{3,7}$,
  C.~Louren\c{c}o$^{3}$,
  J.~Lozano$^{6}$,
  F.~Manso$^{4}$,
  P.~Martins$^{6}$,
  A.~Masoni$^{2}$,
  A.~Neves$^{6}$,
  H.~Ohnishi$^{8}$,
  C.~Oppedisano$^{10}$,
  P.~Parracho$^{3,6}$,
  P.~Pillot$^{11}$,
  T.~Poghosyan$^{12}$,
  G.~Puddu$^{1,2}$,
  E.~Radermacher$^{3}$,
  P.~Ramalhete$^{3,6}$,
  P.~Rosinsky$^{3}$,
  E.~Scomparin$^{10}$,
  J.~Seixas$^{6}$,
  S.~Serci$^{1,2}$,
  R.~Shahoyan$^{3,6}$,
  P.~Sonderegger$^{6}$,
  H.J.~Specht$^{5}$,
  R.~Tieulent$^{11}$,
  G.~Usai$^{1,2}$,
  R.~Veenhof$^{6}$ and
  H.K.~W\"ohri$^{2,6}$\\
  (NA60 Collaboration)}

~

\address{
  $^{~1}$Univ. di Cagliari, Italy;
  $^{~2}$INFN Cagliari, Italy;
  $^{~3}$CERN, Geneva, Switzerland;
  $^{~4}$LPC, Univ.\ Blaise Pascal and CNRS-IN2P3, Clermont-Ferrand, France;
  $^{~5}$Univ.\ Heidelberg, Heidelberg, Germany;
  $^{~6}$IST-CFTP and LIP, Lisbon, Portugal;
  $^{~7}$LLR, Ecole Polytechnique and CNRS-IN2P3, Palaiseau, France;
  $^{~8}$RIKEN, Wako, Saitama, Japan;
  $^{~9}$Univ.\ di Torino, Italy;
  $^{~10}$INFN Torino, Italy;
  $^{~11}$IPN-Lyon, Univ.\ Claude Bernard Lyon-I and CNRS-IN2P3, Lyon, France;
  $^{~12}$YerPhI, Yerevan, Armenia
}

\maketitle\abstracts{The NA60 experiment at the CERN SPS has studied
  low-mass dimuon production in 158A\,GeV In-In collisions. A
  significant excess of pairs is observed above the yield expected
  from neutral meson decays, consistent with a dominant contribution
  from $\pi\pi\to\rho\to\mu\mu$. This paper presents precision results
  on the mass and transverse momentum spectra of the excess pairs.
  The space-time averaged rho spectral function associated to the
  measured mass distribution shows a significant broadening, but
  essentially no mass shift. The slope parameter $T_{eff}$ extracted
  from the spectra rises with dimuon mass up to the $\rho$, followed
  by a sudden decline above. While the initial rise is consistent with
  the expectations for radial flow of a hadronic decay source, the
  decline indicates a transition to an emission source with much smaller
  flow, possibly of partonic origin.}

The ultimate goal of heavy ion collisions is the detection of
signatures of a phase transition from hadronic matter to a deconfined
and/or chirally restored medium. Lepton pairs are a powerful probe of
the hot and dense fireball formed in high-energy nuclear
collisions. They are produced during the entire space-time evolution
of the fireball and freely escape, undisturbed by final state
interactions.  In the invariant mass region $<$1 GeV, thermal
dilepton production is mediated by the broad vector meson $\rho$(770).
In the mass region $>$1 GeV, thermal dileptons may be produced in
either the early partonic or the late hadronic phase of the fireball.

In contrast to real photons, virtual photons decaying into lepton
pairs can be characterized by two variables, mass $M$ and transverse
momentum $p_{T}$.  The mass distribution can be directly connected to
the space-time averaged spectral function of the intermediate vector
meson.  The measurement of $p_{T}$ spectra of lepton pairs may offer
access to their emission region, as $p_{T}$ encodes the key properties
of the expanding fireball (temperature and transverse flow).  While
hadrons always receive the full asymptotic flow reached at the moment
of decoupling from the flowing
medium~\cite{Schnedermann:1993ws,Heinz:2004qz}, lepton pairs are
continuously emitted during the evolution of the system. They are thus
produced with small flow and high temperature at early times, and
larger flow and smaller temperatures at later times.


\begin{figure}[tbp]
\begin{center}
\includegraphics*[width=5.5cm, height=5.5cm]{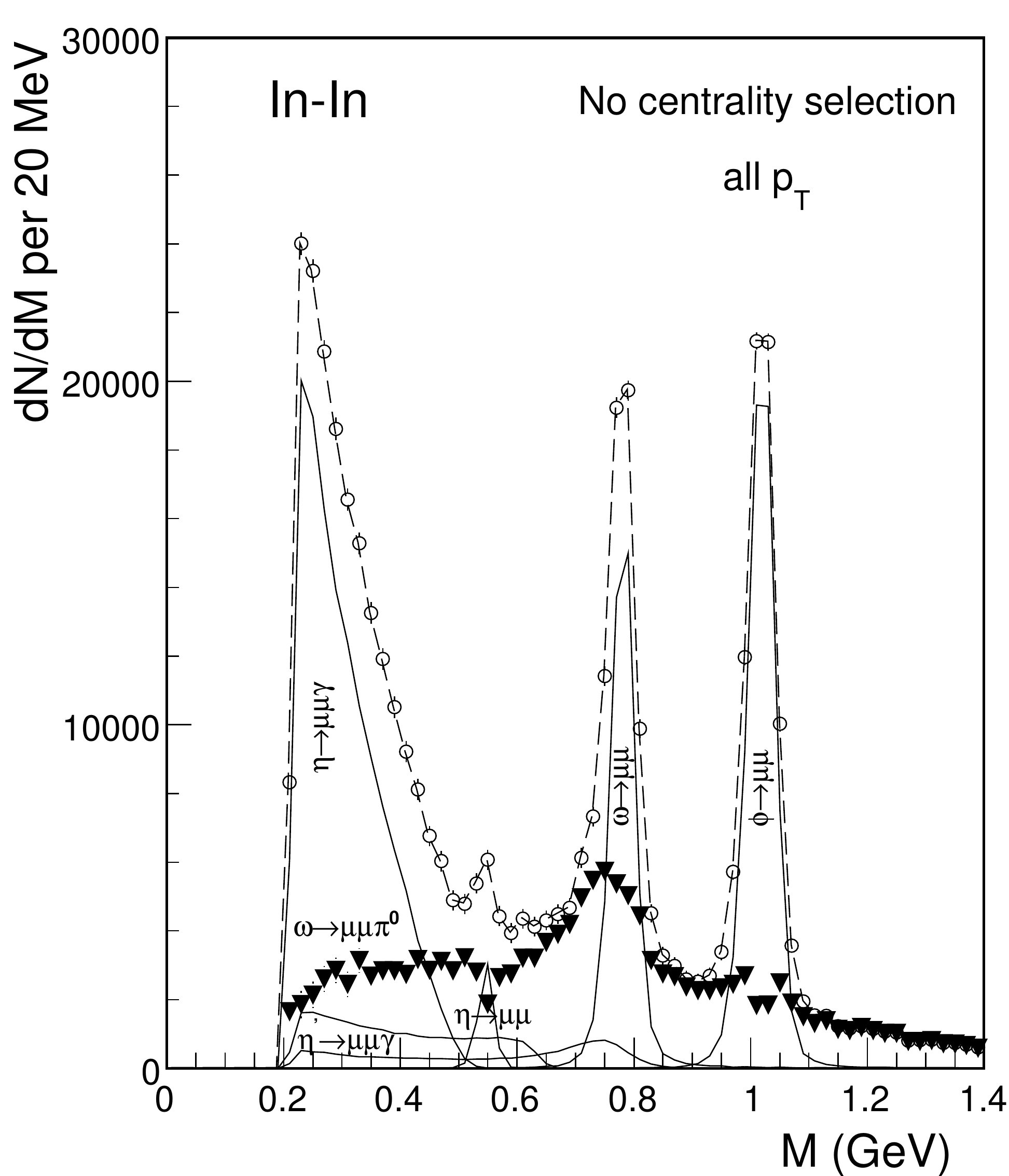}
\hspace{1cm}
\includegraphics*[width=5.5cm, height=5.5cm]{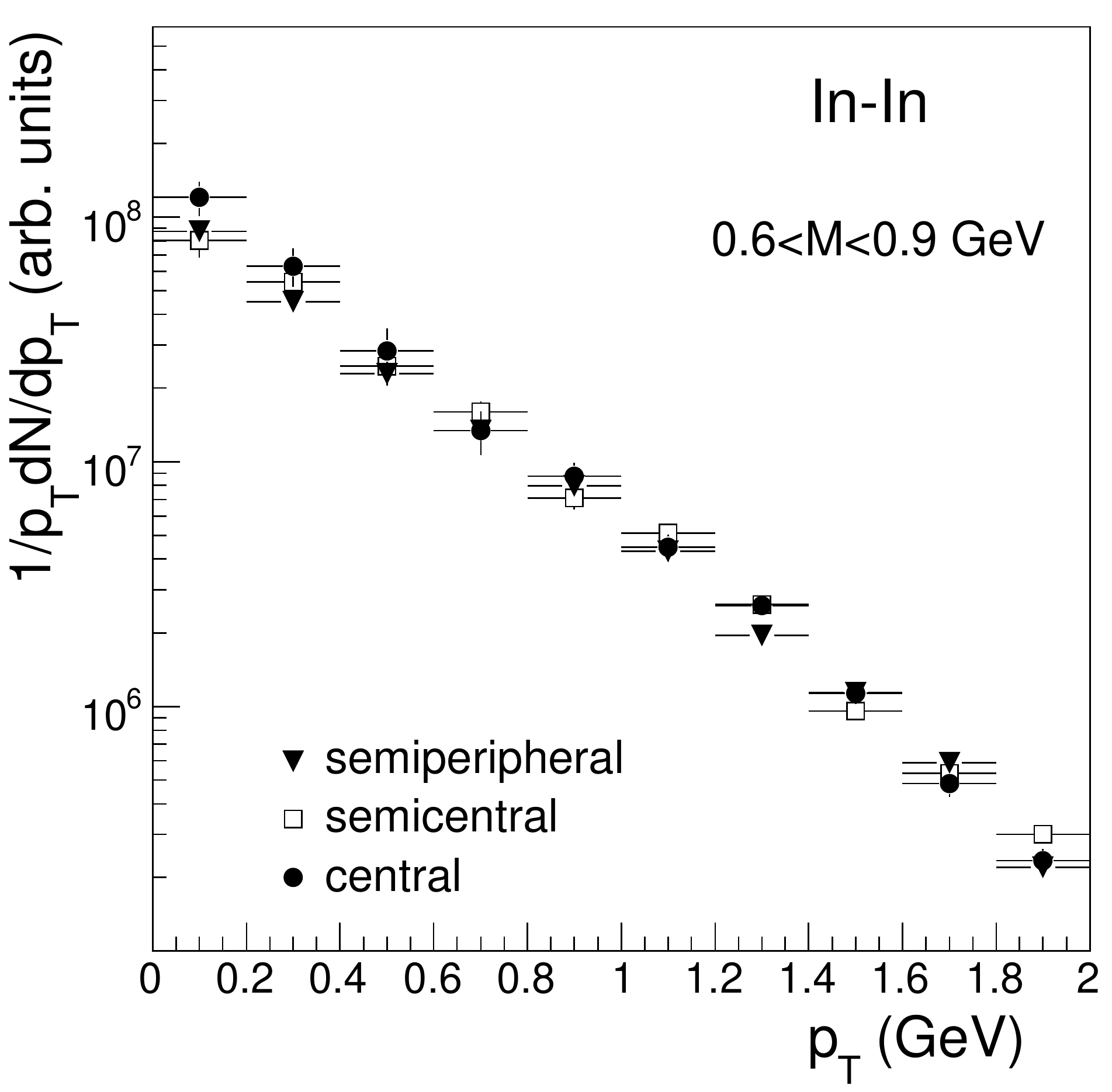} 
\vspace*{-0.3cm}
\caption{Left: Isolation of an excess above the electromagnetic decays
  of neutral mesons (see text). Total data (open circles), individual
  cocktail sources (solid), difference data (thick triangles), sum of
  cocktail sources and difference data (dashed). Open charm not
  subtracted. Right: Transverse momentum spectra of the rho-like mass
  region for three centrality windows, arbitrarily normalized. The
  errors are purely statistical}
   \label{fig1}
\end{center}
\vspace*{-0.8cm}
\end{figure}

The NA60 experiment measured low-mass dimuon production in 158A\,GeV
In-In collisions.  The centrality-integrated net mass spectrum after
subtraction of backgrounds~\cite{Shahoyan:2006eb} is shown in
Fig.~\ref{fig1}. It contains about 430\,000 dimuons in the mass range
$\leq$ 1.4 GeV.  The mass resolution at the $\omega$ is 20 MeV. The
data are divided in centrality bins using the measured charged track
density.  While peripheral data can be described as a superposition of
expected contribution from the electromagnetic decays of neutral
mesons (``hadron cocktail''), this is not possible for the total data,
due to the existence of a strong excess of pairs.  To isolate this
excess without any fits, the decay sources are subtracted from the
data using local criteria which are only based on the measured mass
distribution.  The procedure is illustrated in Fig.~\ref{fig1}. The
narrow $\omega$ and $\phi$ peaks are fixed in order to get a smooth
underlying continuum after subtraction. The yield of the $\eta$
relative to the $\omega$ and $\phi$ is fixed from the data at
p$_{T}$$>$1.0 GeV. This contribution is relevant only for masses
$\leq$0.4 GeV. The $\eta$ two-body and $\omega$ Dalitz decays are then
fixed as well. The ratio $\eta^{'}$/$\eta$ is assumed to be
0.12~\cite{genesis:2003}.
%
%
The $\rho$ is not subtracted.  Open charm, measured to be
0.30$\pm$0.06 of the total yield in the mass interval 1.2$<$$M$$<$1.4
GeV~\cite{Shahoyan:2006qm}, is subtracted throughout (but not
yet in Fig.~\ref{fig1}), with the spectral shape in $M$ and $p_{T}$ as
described by Pythia~\cite{Shahoyan:2006qm}.
For a detailed discussion of the subtraction procedure and associated
systematic uncertainties see Ref.~\cite{Arnaldi:2007ru,Sanja-QM}.
After subtraction of the meson decays and charm, the remaining sample
contains $\sim$150\,000 dimuons. The subtracted data for the $\eta$,
$\omega$ and $\phi$ provide the $p_T$ spectra for these mesons and are
used later for comparison.  A strong excess is found, with respect to
the contribution expected from the ``cocktail'' $\rho$ (bound to the
$\omega$ with $\rho/\omega=1$, as measured in elementary
collisions). The excess is centered around the nominal $\rho$ pole
position and is compatible with a strong broadening of the $\rho$
spectral function and no mass shift. It monotonically increases and
broadens with centrality. More details can be found in
Ref.~\cite{Arnaldi:2006jq,Agakichiev:1997au}

The data are corrected for the acceptance and for the
centrality-dependent reconstruction efficiencies, estimated with an
overlay Monte Carlo simulation (a Monte Carlo dimuon is reconstructed
on top of a real event)~\cite{Damjanovic:2007qm}.
\begin{figure}[t!]
\begin{center}
\includegraphics*[width=5.5cm, height=5.5cm]{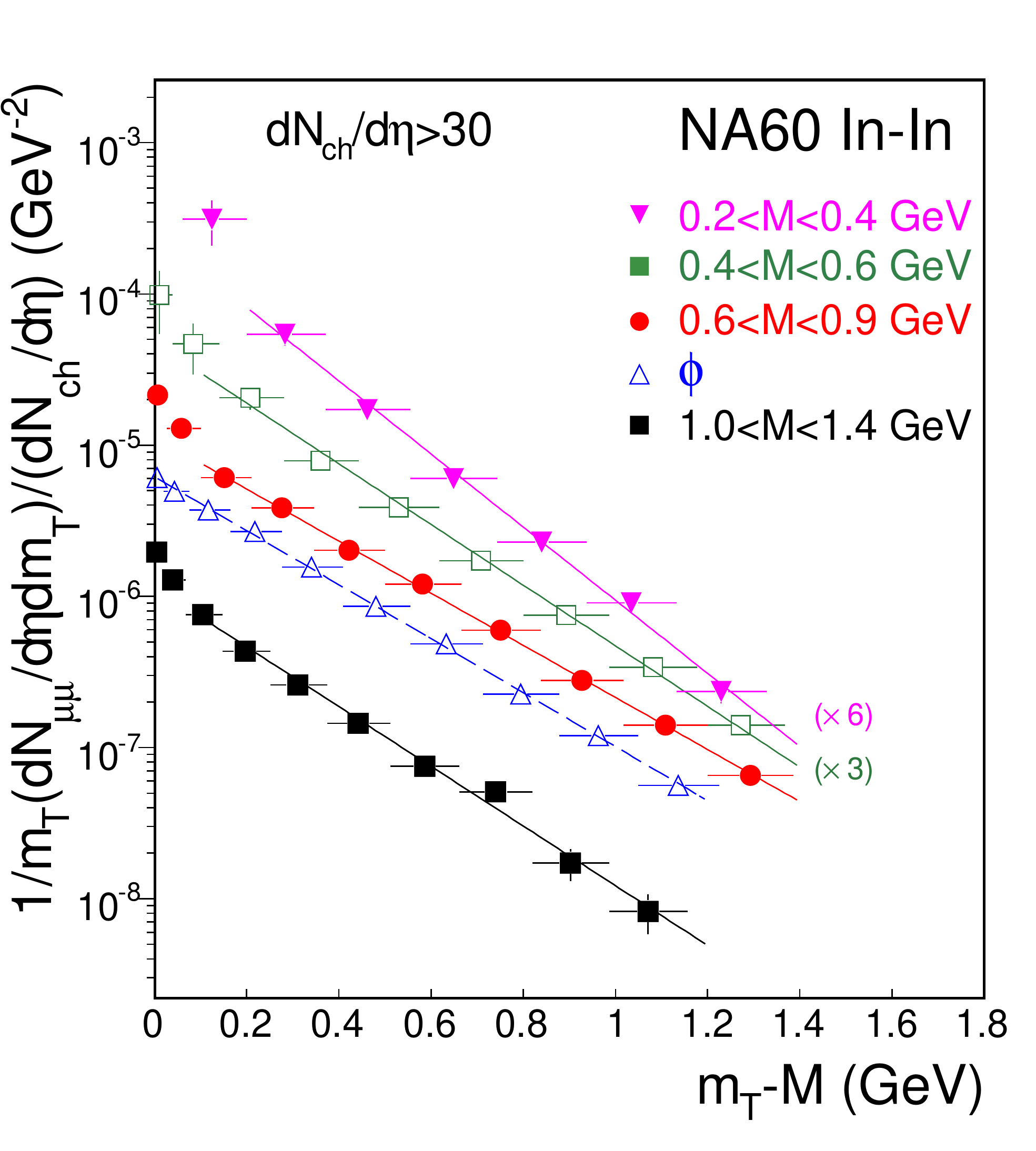}
\hspace{1cm}
\includegraphics*[width=5.5cm, height=5.5cm, clip= 0 2 564 537]{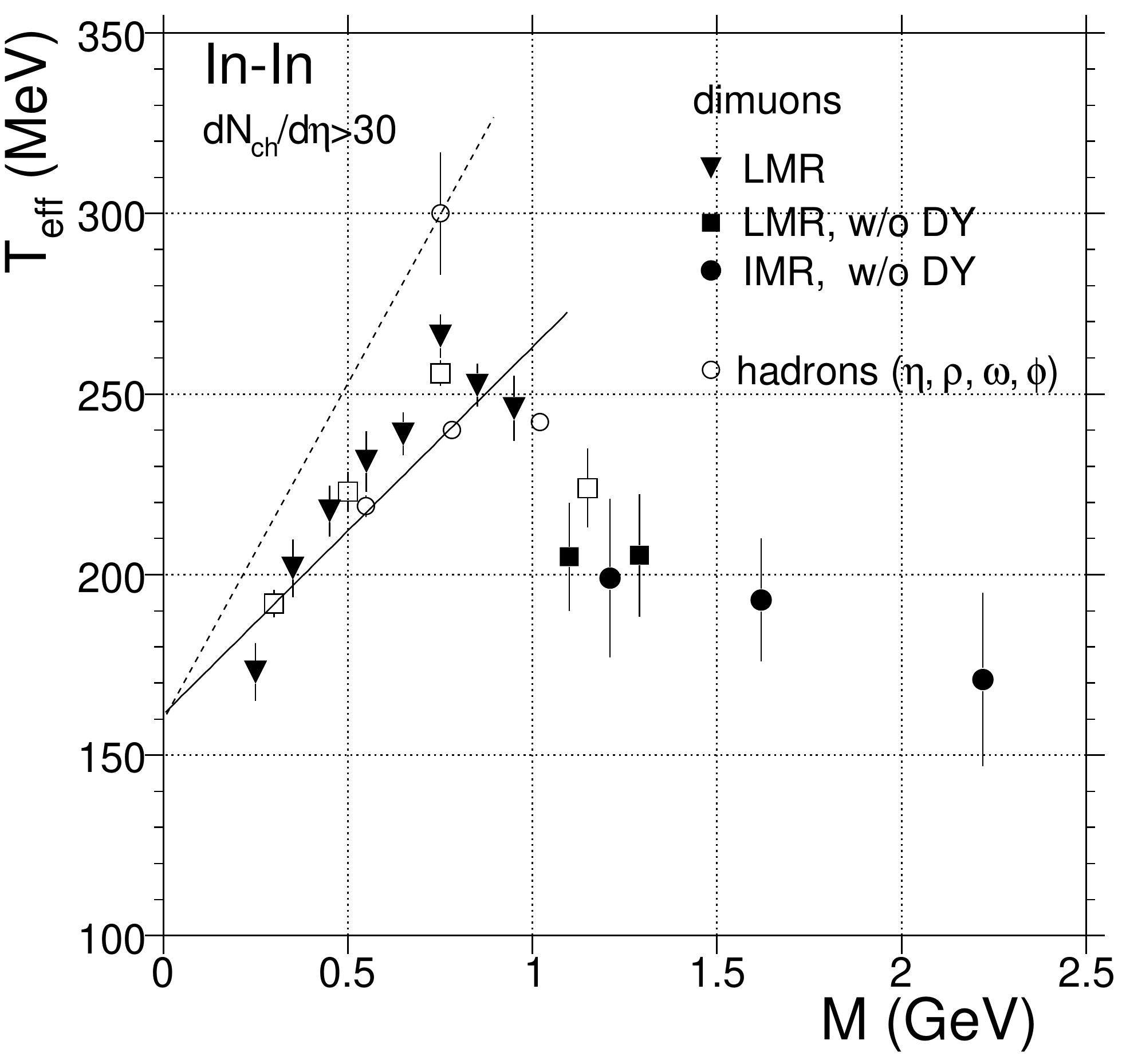}
\vspace*{-0.3cm}
\caption{Left: Transverse mass spectra of the excess for four mass
  windows summed over centralities (excluding the peripheral bin), in
  comparison to the $\phi$ (absolute normalization).  Right: Inverse
  slope parameter $T_\mathrm{eff}$ vs. dimuon mass $M$ for
  $dN_{ch}/d\eta$$>$30. The open squares correspond to the fit lines
  in the left pane. Open charm is subtracted throughout. For
  explanation of the inserted symbols and the errors see text.}
   \label{fig2}
\end{center}
\vspace*{-0.7cm}
\end{figure}
%
%
Results on acceptance-corrected $p_{T}$-spectra for the mass window
0.6$\leq$$M$$\leq$0.9 GeV and non-peripheral collisions are shown in
Fig.~\ref{fig1}; equivalent data for other mass windows can be found
Ref.~\cite{Damjanovic:2007qm}.
The data of the three centrality windows agree within errors;
this also holds for the other mass
windows~\cite{Damjanovic:2007qm}. 
Fig.~\ref{fig2} (left) displays the centrality-integrated data
vs. transverse mass $m_{T}$ for four mass windows; the $\phi$ is
included for comparison.
At very low $m_{T}$, a steepening is observed in all four mass
windows, reminiscent of pion spectra and opposite to the expectation
for radial flow at masses above the pion mass. This steepening is not
observed for the hadron spectra.
The rise only disappears for very peripheral collisions
(4$<$$dN_{ch}/d\eta$$<$10). The lines in Fig.~\ref{fig2} (left) are
fits with the exponential function 1/$m_{T}$ $dN/dm_{T}$ $\propto $
$exp(-m_{T}/T_\mathrm{eff})$. The fits are restricted to the range
0.4$<$$p_{T}$$<$1.8 GeV (roughly 0.1$<$$(m_{T} - M)$$<$1.2 GeV), to
exclude the increased rise at low $m_{T}$.  Fig.~\ref{fig2} (right)
shows $T_\mathrm{eff}$ vs. pair mass, for the mass windows defined in
the left pane (open squares) and for a finer mass binning (closed
triangles).  NA60 results from an independent
analysis~\cite{Shahoyan:2006qm} of the intermediate mass region
(``IMR'') 1.16$<$$M$$<$2.56 GeV, corrected for the contribution from
Drell-Yan pairs, are shown for comparison.  The hadron data as
obtained from the subtraction procedure are also included; the value
for the $\eta$ has been obtained by tuning the GENESIS
code~\cite{genesis:2003} to the $\eta$ Dalitz decay and then referring
back to the required $T_\mathrm{eff}$ of the mother. The errors shown
for the low-mass data (``LMR'') are purely statistical. The systematic
error is smaller than the statistical one (see Ref.~\cite{Sanja-QM}).

The results shown in Fig.~\ref{fig2} (right) can be summarized and
interpreted as follows. The slope parameter $T_\mathrm{eff}$ rises
nearly linearly with mass up to about 270 MeV at the pole position of
the $\rho$, followed by a sudden decline to values of 190-200 MeV for
masses $>$1 GeV. The excess yield in the mass region 0.2$<$$M$$<$0.9
GeV is generally attributed to thermal radiation from the fireball,
dominated by pion annihilation
$\pi^{+}\pi^{-}\rightarrow\rho\rightarrow\mu^{+}\mu^{-}$ via an
in-medium modified $\rho$~\cite{Arnaldi:2006jq}. The NA60 data in this
region are now nearly quantitatively described by theoretical
models~\cite{vanHees:2006ng,Ruppert:2006hf,Dusling:2007rh}. The linear
rise of $T_\mathrm{eff}$ with $M$ over the whole region up to the
$\rho$ peak is reminiscent of radial flow of a hadronic source.  The
hadron data show a similar linear rise, but their absolute values are
surprisingly close to the excess values, contrary to the expectation
for the temperature-flow folding. Hadrons should have larger flow,
being emitted at freeze-out, while leptons are emitted throughout the
whole lifetime of the fireball. To solve this seeming contradiction
the vacuum $\rho$ was studied. It can be isolated disentangling the
peak from the broad continuum (Fig.~\ref{fig1},
left)~\cite{Damjanovic:2006bd,Damjanovic:2007qm,vanHees:2006ng,Ruppert:2006hf,Dusling:2007rh}.
It is found $T_\mathrm{eff}$=300$\pm$17 MeV for the peak and 231$\pm$7
for the underlying continuum in the window 0.6$<$$M$$<$0.9 MeV. The
subtraction of a $\rho$ modelled with the measured $T_\mathrm{eff}$
from the continuum data leads to differences of 4$-$20 MeV in
$T_\mathrm{eff}$ depending on the closeness to the pole.
The following interpretation emerges: the $\rho$ is maximally coupled
to pions and it is thus representative of the true freeze-out
parameters of the fireball. The other hadrons freeze-out earlier, due
to their smaller coupling to the pions. The large gap in
$T_\mathrm{eff}$ between the vacuum $\rho$ and the excess points is
consistent with the expectations for the temperature-flow folding of
the leptons. The linear rise of $T_\mathrm{eff}$ with $M$ is
consistent with the expectations for radial flow of a hadronic source
($\pi\pi \rightarrow \rho$) decaying into lepton pairs.
Theoretical understanding of these results is
underway~\cite{vanHees:2006ng,Ruppert:2006hf,Dusling:2007rh}, but does
not yet describe the data in a satisfactory way.  The large gap ($>$50
MeV) in $T_\mathrm{eff}$ between the vacuum $\rho$ and the $\omega$
only disappears for the lowest peripheral ``pp-like'' selection
4$<$$dN_{ch}/d\eta$$<$10, with $T_\mathrm{eff}$=198$\pm$6 MeV for the
$\rho$ and 201$\pm$4 MeV for the $\omega$. This implies that the ``hot
$\rho$'' is intimately connected to pions, disappearing as the
$\pi\pi$ contribution to $\rho$ production vanishes (with only the
``cocktail'' $\rho$ left).  The sudden decline of $T_\mathrm{eff}$ at
masses $>$1 GeV is the other most remarkable feature of the present
data, which may indicate a transition to a source of a different
nature. Extrapolating the lower-mass trend to beyond 1 GeV, a jump by
about 50 MeV down to a low-flow situation is extremely hard to
reconcile with emission sources of dominantly hadronic origin. If the
rise is due to flow, then the sudden loss of flow may be explained as
a transition to a qualitatively different source, suggesting
dominantly early (partonic) processes like
$q\bar{q}\rightarrow\mu^{+}\mu^{-}$ for which flow has not yet built
up~\cite{Ruppert:2006hf}. While still
controversial~\cite{vanHees:2006ng}, this may well represent the first
direct evidence for thermal radiation of partonic origin.

In conclusion, we have found strong evidence for radial flow in the
region of thermal dilepton emission which has previously been
associated with the $\rho$ spectral function. The transition to a
low-flow region above may signal a transition from a hadronic to a
partonic source.


We are grateful to H.~van~Hees, R.~Rapp, T.~Renk and J.~Ruppert for
useful discussions. We acknowledge support from the BMBF (Heidelberg
group), the C. Gulbenkian Foundation, Lisbon, and the Swiss Fund
Kidagan (YerPhi group) and FCT (Lisbon group). M. Floris
acknowledges support from the European Union ``Marie Curie'' Programme
for the participation at the ``XLIII Rencontres de Moriond''
Conference.


\end{document}